\documentclass[pra,twocolumn,superscriptaddress,10pt,noshowpacs]{revtex4}
\usepackage[english]{babel}
\usepackage[T1]{fontenc}
\usepackage[utf8]{inputenc}
\usepackage{graphicx,epstopdf}
\usepackage{amsmath}

\usepackage{amsfonts}
\usepackage{bbm}
\usepackage{amssymb}
\usepackage{color}
\usepackage{latexsym}
\usepackage{caption}
\usepackage{subcaption}
\usepackage{times,txfonts}
\usepackage{hyperref}
\hypersetup{
    colorlinks=true,
    citecolor=blue,
    filecolor=green,
    linkcolor=blue,
    urlcolor=red,
}

\begin{document}

\title{Electronic states in a bilayer graphene quantum ripple}

\author{M. C. Ara\'{u}jo}
\email{michelangelo@fisica.ufc.br}
\affiliation{Universidade Federal do Cear\'a (UFC), Departamento de F\'isica,\\ Campus do Pici, Fortaleza - CE, C.P. 6030, 60455-760 - Brazil.}

\author{A. C. A. Ramos}
\email{antonio.ramos@ufca.edu.br}
\affiliation{Centro de Ci\^{e}ncias e Tecnologias, Universidade Federal do Cariri, 63048-080, Juazeiro do Norte, Brazil}

\author{J. Furtado}
\email{job.furtado@ufca.edu.br}
\affiliation{Centro de Ci\^{e}ncias e Tecnologias, Universidade Federal do Cariri, 63048-080, Juazeiro do Norte, Brazil}

\date{\today}


\begin{abstract}

In this paper, we investigate the influence of the geometry in the electronic states of a quantum ripple surface. We have considered an electron governed by the spinless stationary Schr\"{o}dinger equation constrained to move on the ripple surface due to a confining potential from which the Da Costa potential emerges. We investigate the role played by the geometry and orbital angular momentum on the electronic states of the system.  

\end{abstract}

\maketitle


\section{Introduction}

As reported in transmission electron microscopy (TEM) \cite{meyer2007structure,meyer2007roughness} and scanning tunneling microscopy experiments (STM) \cite{ishigami2007atomic,stolyarova2007high}, suspended graphene samples tend to exhibit seemingly random spontaneous curvatures which can be visualized as ripples ranging from a few angstroms in height to several nanometers in length. Curvature effects in graphene can arise due to disorder, i.e., imperfections or irregularities in the material’s structure caused by lattice defects, impurities, or topographical variations on its surface. Disorder can significantly affect the electronic properties of a low-dimensional material, and this has been extensively investigated for graphene \cite{ludwig1994integer,Nersesyan:1994zz,Nersesyan:1995cdf,stauber2005disorder,Peres:2006zz}. Inspired by the physics of nanotubes and fullerenes, many studies on the electronic properties of curved graphene address curvature induced by topological defects \cite{tamura1994disclinations,charlier2001electronic,Gonzalez:2000ovj,cortijo2007electronic,Cortijo:2006xs}, where it has been found that characteristic charge anisotropies can arise due to conical defects on an average flat surface, which could be observed in STM or TEM experiments. Such anisotropies were later observed in experiments with scanning single-electron transistors \cite{martin2008observation} and electrostatic force microscopy \cite{lu2006electrostatic}.

Indeed, the geometry of graphene plays a central role in the material’s electronic structure. The presence of curvature on the surface is associated with fascinating effects, such as the formation of quantized energy levels, the emergence of geometric phases, and curvature-induced forces that influence the dynamics and trajectory of particles \cite{costa, costa1}, chiral properties \cite{dandoloff2004quantum,atanasov2009geometry,atanasov2015helicoidal}, the existence of pseudo-magnetic fields \cite{guinea2010generating} and charge inhomogeneity \cite{de2007charge}, which can arise in corrugated \cite{atanasov2010tuning} and rippled \cite{de2007charge} layers.

A continuum limit approach, which allows for the formulation of the Hamiltonian of an electron on curved surfaces, is feasible through the thin-layer squeezing method \cite{jensen1971quantum}. Starting from a three-dimensional Schrödinger Hamiltonian, this method yields an attractive geometric potential that depends on both the mean curvature and the Gaussian curvature, known as the da Costa potential \cite{costa,costa1,matsutani1992berry,daSilva:2016wxz}. This approach can also be extended to include external fields and spin, as noted in references \cite{ferrari2008schrodinger,wang2014pauli,burgess1993fermions}. Among the studied surfaces, we highlight the torus, as carbon nanotori are present in nanoelectronics, biosensors, and quantum computing \cite{Goldsmith, Goldsmith2}. In Ref. \cite{encinosa}, for instance, bound states and their respective curvature-induced eigenvalues were calculated for an electron governed by the Schrödinger equation. A similar analysis was conducted in Ref. \cite{GomesSilva:2020fxo} with the inclusion of external fields in the system. Additionally, governed by the Pauli equation, a spin-1/2 charged particle confined to the surface of a torus was investigated in Ref. \cite{AGM}, and analytical solutions for the Dirac equation in (2+1) dimensions were found in Ref. \cite{Yesiltas:2018zoy} for cases where the Fermi velocity is constant and position-dependent. An extension of this latter work to include external fields was carried out in Ref. \cite{Yesiltas:2021crm}. More recently, a potential qubit encoding in the energy levels of a graphene nanotoroid was investigated in Ref. \cite{Furtado:2022uvk}. 

Given that the electronic properties of two-dimensional carbon materials, such as graphene \cite{katsnelson, geim, castro} and phosphorene \cite{phosphorene}, have been shown to be highly dependent on sample geometry \cite{CostaFilho:2020sbw, Aguiar:2020dgi}, these configurations can be used as highly relevant analog models for high-energy physics \cite{Capozziello:2020ncr, Cvetic:2012vg, Pourhassan:2018wjg, Acquaviva:2022yiq, Iorio:2013ifa, Iorio:2010pv, Iorio:2014pwa}. Here, we highlight the catenoid geometry, which is configured as a minimal surface equivalent to that of a wormhole \cite{dandoloff}. For example, in Refs. \cite{gonzalez} and \cite{, pincak}, the authors proposed a nanotube bridge connecting a graphene bilayer. The electronic properties of such structures were described in Ref. \cite{novaes2010electronic}, where it was reported that for metallic nanotubes, conductance is approximately independent of their length but highly dependent on the nanotube-bilayer junction. The exact opposite occurs for semiconductor nanotubes, where conductance becomes dependent on the nanotube length but independent of the junction. In Refs. \cite{dandoloff} and \cite{dandoloff2}, to achieve a smooth connection between the bridge and the asymptotically flat surfaces, a single catenoid surface was proposed. The effects of geometry and external electric and magnetic fields on the electronic properties of a graphene catenoid structure were investigated in Ref. \cite{euclides}. Additionally, electronic states of a graphene bilayer connected by a generalized Ellis-Bronnikov wormhole-type bridge were investigated in Ref. \cite{deSouza:2022ioq}.

In addition to the geometric potential discussed above (third paragraph), it is expected that curvature may also produce position-dependent mass (PDM) effects, as the homogeneity of the lattice is broken. Although PDM effects have been extensively studied on flat surfaces \cite{pdm1,pdm2,pdm3,pdm4}, only recently has an extension of the da Costa’s method for treatment on curved surfaces been proposed, see Ref. \cite{moraes}. PDM considerations for an electron confined to a catenoid were addressed in Refs. \cite{Yesiltas:2021dpm} and \cite{silva2021position}.

In this paper, we investigate the influence of geometry on the electronic states of a quantum ripple surface. We consider an electron governed by the spinless stationary Schrödinger equation, constrained to move on a Gaussian-type ripple due to a confining potential from which the da Costa potential emerges. We examine the roles played by both geometry and orbital angular momentum on the electronic states of the system.  

This paper is organized as follows: In Section \ref{section2}, we present the geometry and dynamics of an electron governed by the Schrödinger equation with the da Costa potential, resulting from the thin-layer squeezing method. The main characteristics of this potential are then discussed. Using the method of separation of variables, we derive a one-dimensional equation describing a position-dependent mass system in the presence of an effective potential that depends on the orbital quantum number $\ell$. A qualitative analysis of this potential is also performed. In Section \ref{section333}, we investigate how the geometric parameters characterizing the Gaussian-type ripple surface affect the bound states of the system. Finally, in Section \ref{section444}, we present our conclusions.


\section{Electron on a Gaussian ripple surface}
\label{section2}

\begin{figure*}[ht]
    \centering
    \includegraphics[scale=0.5]{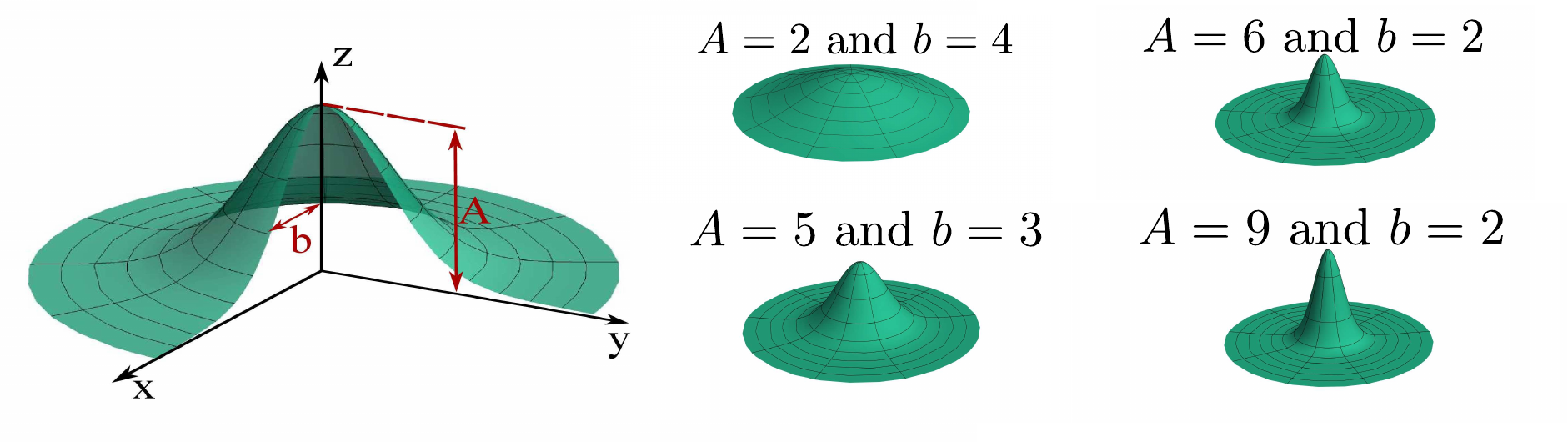}
    \caption{Quantum ripple coordinate system and some configurations}
    \label{fig1}
\end{figure*}

In this section, we introduce the geometry and dynamics of an electron on a quantum Gaussian-like ripple surface, considering the effects of the system curvature and geometry. We consider an electron described by a quadratic dispersion relation, employing the Schrödinger equation on curved surfaces to properly describe the system. As shown in Fig. \ref{fig1}, the ripple is realized as a generated Gaussian surface. Our coordinate system is also depicted in Fig. \ref{fig1}.

After squeezing the electron wave-function on the surface, the spinless stationary Schr\"{o}dinger equation has the form
\begin{equation}
\label{constantmassschrodinger}
    -\frac{\hbar^2}{2m^*}\nabla^2 \Psi +V_{dc}\Psi=E\Psi,
\end{equation}
where $\nabla^{2}\Psi=\frac{1}{\sqrt{g}}\partial_{a}(\sqrt{g}g^{ab}\partial_{b}\Psi)$ is the Laplacian operator in curved spaces, $g^{ab}$ is the induced metric, $m^*$ is the effective mass of the electron and $V_{dC} = - \frac{\hbar^2}{2m^{*}}(H^2-K)$ is the da Costa potential \cite{costa}. Note the dependence of this latter on both mean curvature $H$ and Gaussian curvature $K$. 


The parameterization we adopted for the surface shown in Fig. \ref{fig1} is 
\begin{eqnarray}
\label{surface}
\nonumber \vec{x}(r,\phi)&=&r\cos\phi \, \hat{i} + r\sin\phi \, \hat{j}+ A\,  e^{-r^2/b^2} \, \hat{k},
\end{eqnarray}
where $A\geq 0$ is the height of the Gaussian bump and $b>0$ is the standard deviation, geometrically represented by the width at half-height. The coordinates $r\in(0,\infty)$ and $\phi\in[0,2\pi)$ cover the whole ripple surface. It is important to highlight that for a real two-dimensional material, some cutoffs in the coordinates must be imposed in order to properly guarantee the validity of the continuum limit here employed. For a graphene-like material, for example, the bonds in the lattice are around $1,43 \AA$ \cite{graphene-bond}. Thus, we must assume $r>>1,43 \AA$ in order to ensure the validity of the continuum limit approach.

\begin{figure*}
    \centering
    \includegraphics[scale=0.4]{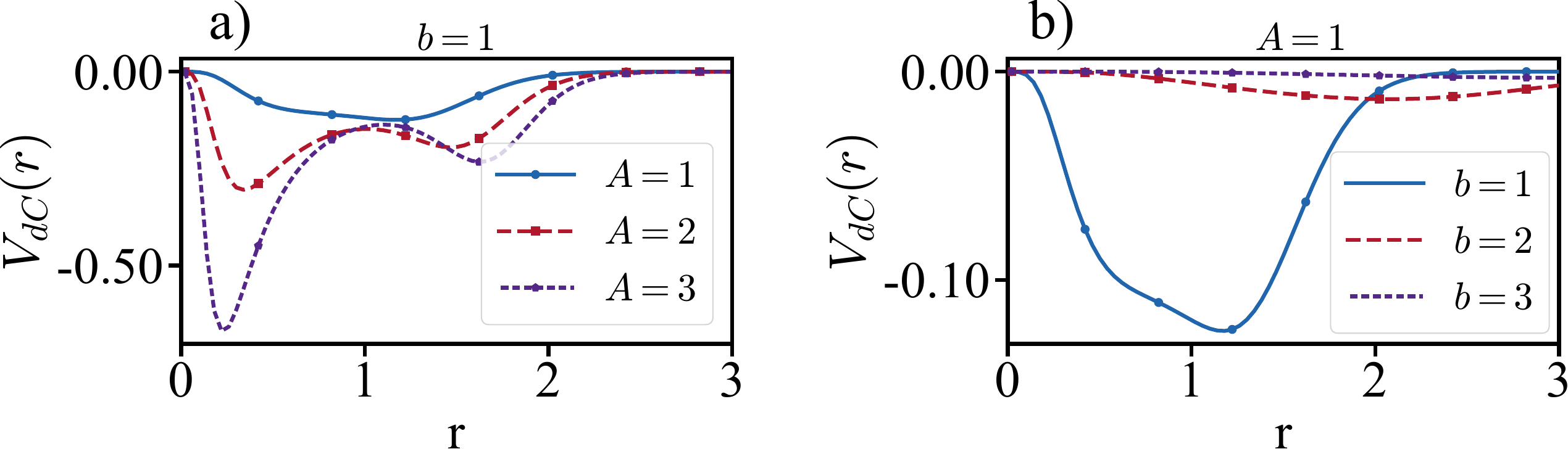}
    \caption{Plots of the da Costa potencial for the quantum Gaussian ripple. In the left panel we have considered $b=1$ and three values of $A$, namely $A=1$, $A=2$ and $A=3$. In the right panel we have considered $A=1$ and three values of $b$, namely, $b=1$, $b=2$ and $b=3$.}
    \label{Fig11}
\end{figure*}

In the chosen coordinate system, the interval reads
\begin{eqnarray}\label{dsdoisfrdr2plusr2dphi2}
ds^{2}=f(r)dr^2+r^2d\phi^2,
\end{eqnarray}
with
\begin{equation}
    f(r)=1+4\left(\frac{A^2}{b^2}\right)\left(\frac{r^2}{b^2}\right)e^{-2r^2/b^2}.
\end{equation}
The components of the diagonal induced metric tensor are 
\begin{eqnarray}
    g_{rr}&=&f(r);\\
    g_{\phi\phi}&=&r^2,
\end{eqnarray}
and the nonvanishing components of the Christoffel symbols can be calculated as
\begin{eqnarray}
    \Gamma^{r}_{rr}&=&\frac{4 A^2 r(b^2-2r^2)}{b^6e^{2r^2/b^2}+4A^2b^2r^2};\\
    \Gamma^{r}_{\phi\phi}&=&-\frac{b^4r^3}{b^4r^2+4A^2r^4e^{-2r^2/b^2}};\\
    \Gamma^{\phi}_{r\phi}&=&\Gamma^{\phi}_{\phi r}=\frac{1}{r}.
\end{eqnarray} 
The da Costa potential, in turn, is given by
\begin{eqnarray}\label{dacostapot}
   V_{dC}(r)=-\frac{2 \hbar^2 r^4 \left(2 A^3+A b^2 e^{\frac{2 r^2}{b^2}}\right)^2}{m^{\ast} \left(4 A^2 r^2+b^4 e^{\frac{2 r^2}{b^2}}\right)^3}.
\end{eqnarray} From \eqref{dacostapot}, we see that the da Costa potential is entirely negative, since $A$ and $b$ are non-negative parameters. In Fig. \ref{Fig11}, we have depicted $V_{dC}(r)$ in order to obtain a better understanding of the system itself. As a general behaviour, we must notice that the da Costa potential vanishes both in $r=0$ as in the limit $r\rightarrow\infty$, i.e., $lim_{r\rightarrow\infty}V_{dC}=0$. In Fig. \ref{Fig11}a we have considered $b=1$ and three values of $A$, namely $A=1$, $A=2$ and $A=3$. As we can see, $V_{dC}(r)$ exhibits a global minimum whose depth increases with the increase of the parameter $A$. On the other hand, in Fig. \ref{Fig11}b, we have considered $A=1$ and three values of $b$, namely, $b=1$, $b=2$ and $b=3$. From that, we can see that the global minimum decreases with the increase of $b$. It is important to highlight that in Ref. \cite{Liu}, the authors find that there is no geometric potential for a Dirac fermion on a two-dimensional curved surface of revolution. However, in this work, we consider a quadratic dispersion relation for the charge carriers, which means they are described by the Schrödinger equation instead of the Dirac equation. Therefore, we must consider the da Costa’s potential.

The axial symmetry leads to a periodic behavior of the wave function in the form
\begin{equation}
\label{axialsymmetry}
\Psi(r,\phi)=\psi(r)e^{i \ell\phi},
\end{equation}
where $\ell$ is the orbital quantum number. Substituting Eq. \eqref{axialsymmetry} into Eq. \eqref{constantmassschrodinger}, we can write the stationary Schr\"{o}dinger equation as
\begin{eqnarray}
\label{freenonhermitianequation}
\Sigma_1(r)\frac{d^2\psi(r)}{dr^2}+\Sigma_2(r)\frac{d\psi(r)}{dr}+\Sigma_3(r)\psi(r)=E\psi(r),
\end{eqnarray}
with $\Sigma_1(r)$, $\Sigma_2(r)$ and $\Sigma_3(r)$ being functions of the coordinate $r$ given respectively by:
\begin{widetext}
\begin{eqnarray}
    \Sigma_1(r)&=&-\frac{\hbar^2 }{2 m^{\ast} \left(1+\frac{4 A^2 r^2 e^{-\frac{2 r^2}{b^2}}}{b^4}\right)},\\
    \Sigma_2(r)&=&-\frac{b^2 \hbar^2 e^{\frac{2 r^2}{b^2}} \left(8 A^2 r^4+b^6 e^{\frac{2 r^2}{b^2}}\right)}{2 m^{\ast} r \left(4 A^2 r^2+b^4 e^{\frac{2 r^2}{b^2}}\right)^2},\\
    \Sigma_3(r)&=&\frac{\hbar^2 \left(-16 A^6 r^6-16 A^4 b^2 r^6 e^{\frac{2 r^2}{b^2}}+\ell^2 \left(4 A^2 r^2+b^4 e^{\frac{2 r^2}{b^2}}\right)^3-4 A^2 b^4 r^6 e^{\frac{4 r^2}{b^2}}\right)}{2 m^{\ast} r^2 \left(4 A^2 r^2+b^4 e^{\frac{2 R^2}{b^2}}\right)^3}.
\end{eqnarray}
\end{widetext}

\begin{figure*}[ht!]
    \centering
    \includegraphics[scale=0.4]{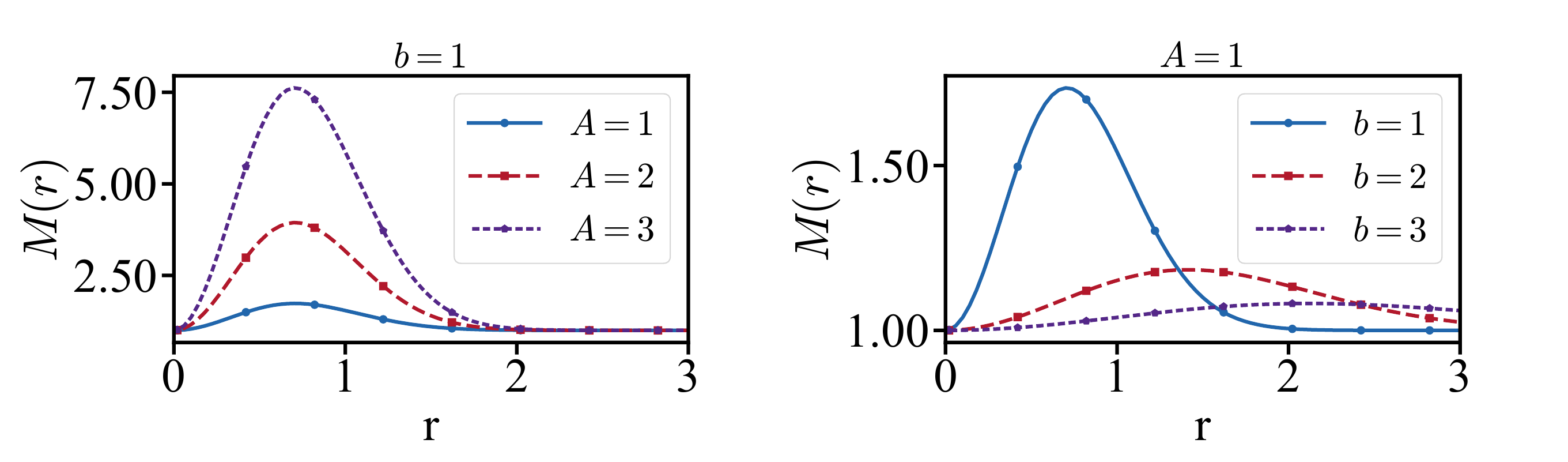}
    \caption{Plots of the effective mass of the electron on the quantum Gaussian ripple. In the left panel we have considered $b=1$ and three values of $A$, namely $A=1$, $A=2$ and $A=3$. In the right panel we have considered $A=1$ and three values of $b$, namely, $b=1$, $b=2$ and $b=3$.}
    \label{Fig3}
\end{figure*} 

In principle, the presence of the first-order derivative in the above equation leads to a non-Hermitian Hamiltonian operator. However, due to the symmetry of the geometric surface considered here, Eq. \eqref{freenonhermitianequation} is manifestly invariant under parity transformations. Since we are considering a time-independent version of the Schrödinger equation (see Eq. \eqref{constantmassschrodinger}), we can conclude that our system also exhibits invariance under time-reversal transformations. Therefore, as long as the space-time reflection symmetries are preserved, the eigenvalue spectrum of the system is expected to be real \cite{Bender, Bender2}, making the aforementioned non-Hermiticity not an issue. Furthermore, there exists an equivalent Hermitian Hamiltonian by a simple change of variable. Let

\begin{equation}
    \psi(r)=\exp \left(\frac{1}{4} \left(\log \left(4 A^2 r^2+b^4 e^{\frac{2 r^2}{b^2}}\right)-\frac{2 r^2}{b^2}-2 \log (r)\right)\right)y(r),
\end{equation} then 
\begin{equation}\label{modifiedschrodingerequation}
    \Sigma_1(r)\frac{d^2y(r)}{dr^2} + V_{eff}(r)y(r)=Ey(r),
\end{equation} is the new Hermitian Schrödinger-type equation in the presence of an effective potential $V_{eff}$ which depends on the orbital quantum number $\ell$, the coordinate $r$, and the geometric parameters of the ripple. Here,
\begin{widetext}
\begin{eqnarray}
    V_{eff}(r)\equiv\frac{h^2 \left(64 A^6 \left(4 \ell^2-1\right) r^6+64 A^4 r^4 e^{\frac{2 r^2}{b^2}} \left(3 b^4 \ell^2+r^4\right)+8 A^2 b^4 r^2 e^{\frac{4 r^2}{b^2}} \left(b^4 \left(6 \ell^2-3\right)+12 b^2 r^2-10 r^4\right)+b^{12} \left(4 \ell^2-1\right) e^{\frac{6 r^2}{b^2}}\right)}{8 m^{*} r^2 \left(4 A^2 r^2+b^4 e^{\frac{2 r^2}{b^2}}\right)^3}.\nonumber\\
\end{eqnarray}       
\end{widetext} Note that the quadratic dependence on $\ell$ ensures that there is no chirality breaking in the system, which is an important feature to be mentioned.

In order to write Eq. \eqref{modifiedschrodingerequation} in an even more convenient way, let $y(r)=s(r)X(r)$. Now, assuming
\begin{equation}
    s(r)=\exp \left(-4 A^2 \left(\frac{\log \left(4 A^2 r^2+b^4 e^{\frac{2 r^2}{b^2}}\right)}{8 A^2}-\frac{r^2}{4 A^2 b^2}\right)\right), 
\end{equation} we can write
\begin{equation}\label{eq_pdm}
    -\frac{\hbar^2}{2}\frac{d}{dr}\left[\frac{\bar{\Sigma}_1(r)}{m^*}\frac{dX(r)}{dr}\right]+\Bar{V}_{eff}(r)X(r)=EX(r),
\end{equation}
where 
\begin{equation}\label{effective_pot}
    \Bar{V}_{eff}(r)=V_{eff}(r)-\frac{\hbar^2}{2}\frac{\bar{\Sigma}_1(r)}{m^*}\left[\frac{s''(r)}{s(r)}\right],
\end{equation} is defined to be a new effective potential with $\bar{\Sigma}_1(r) = (- 2 m^*/\hbar^2)\, \Sigma_1(r)$. In this new format, Eq. \eqref{eq_pdm} can be identified as the Schrödinger equation governing a system with a position-dependent mass \cite{pnbilayer, sinner, pdm1, pdm2, pdm3, pdm4, moraes}, namely
\begin{equation}
    M(r)=\frac{m^*}{\bar{\Sigma}_1(r)}.
\end{equation} In Fig. \ref{Fig3}a, we have depicted the general behaviour of the effective mass $M(r)$ considering $b=1$ and three values of $A$ ($A=1, 2, \text{ and } 3$). As we can see, the effective mass always reaches its maximum at the same point, regardless of the value assumed by parameter $A$. In Fig. \ref{Fig3}b, on the other hand, we considered A=1 and three values of b ($b=1,2, \text{ and } 3$). Note that now the maximum shifts to the right as the value of b increases. In fact, it is straightforward to show that $M(r)$  reaches its maximum at $r_{max}=b/\sqrt{2}$.  

\begin{figure*}
    \centering
    \includegraphics[scale=0.4]{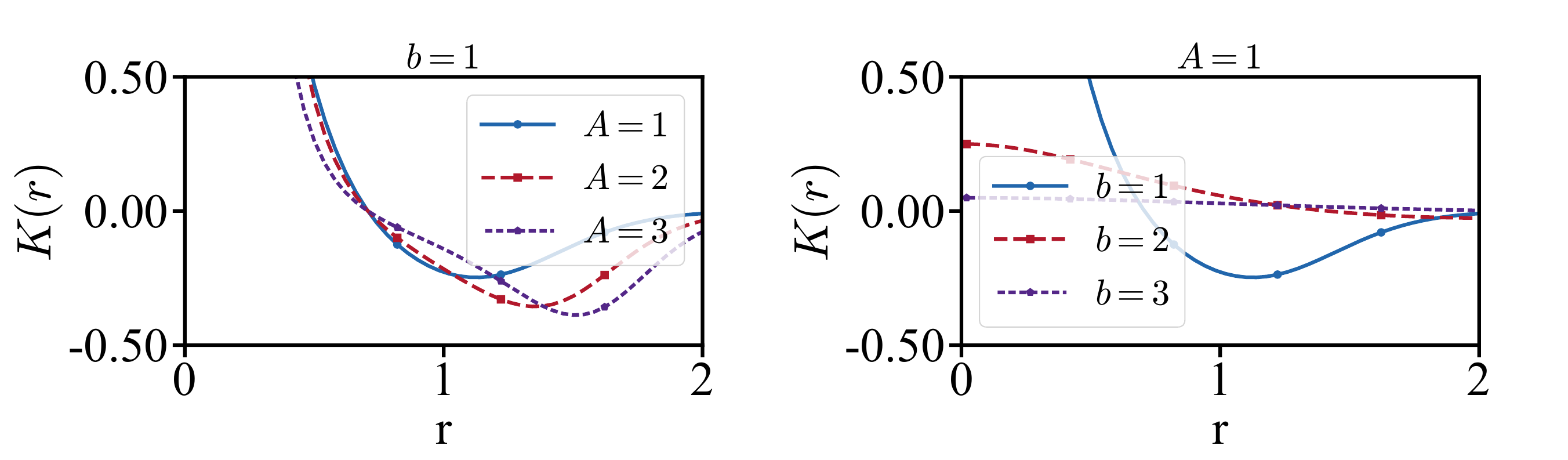}
    \caption{Gaussian curvature for the Gaussian-like ripple. In the left panel we have considered $b=1$ and three values of $A$, namely $A=1$, $A=2$ and $A=3$. In the right panel we have considered $A=1$ and three values of $b$, namely, $b=1$, $b=2$ and $b=3$.}
    \label{fig5}
\end{figure*}

At this point, it is convenient to discuss a few aspects of the Gaussian curvature $K$ in our problem. It follows from Eq. \eqref{dsdoisfrdr2plusr2dphi2} that
\begin{equation}
    K(r)=\frac{4 A^2 b^2 e^{\frac{2 r^2}{b^2}} \left(b^2-2 r^2\right)}{\left(4 A^2 r^2+b^4 e^{\frac{2 r^2}{b^2}}\right)^2},
\end{equation} from which we see that the Gaussian curvature exhibits regions with negative and positive signs and a single point where it vanishes, see Fig. \ref{fig5}. Such a null value point occurs at $r=b/\sqrt{2}$, which is precisely where $M(r)$ reaches its maximum. This can be seen as an indication that the effective mass can be expressed in terms of the Gaussian curvature, which is entirely true since

\begin{equation}
    M(r)=\frac{2 A m^*}{b^3 e^{r^2/b^2}}\sqrt{\frac{b^2-2r^2}{K(r)}}.
\end{equation}


\subsection{Qualitative analysis}

\begin{figure*}
    \centering
    \includegraphics[scale=0.4]{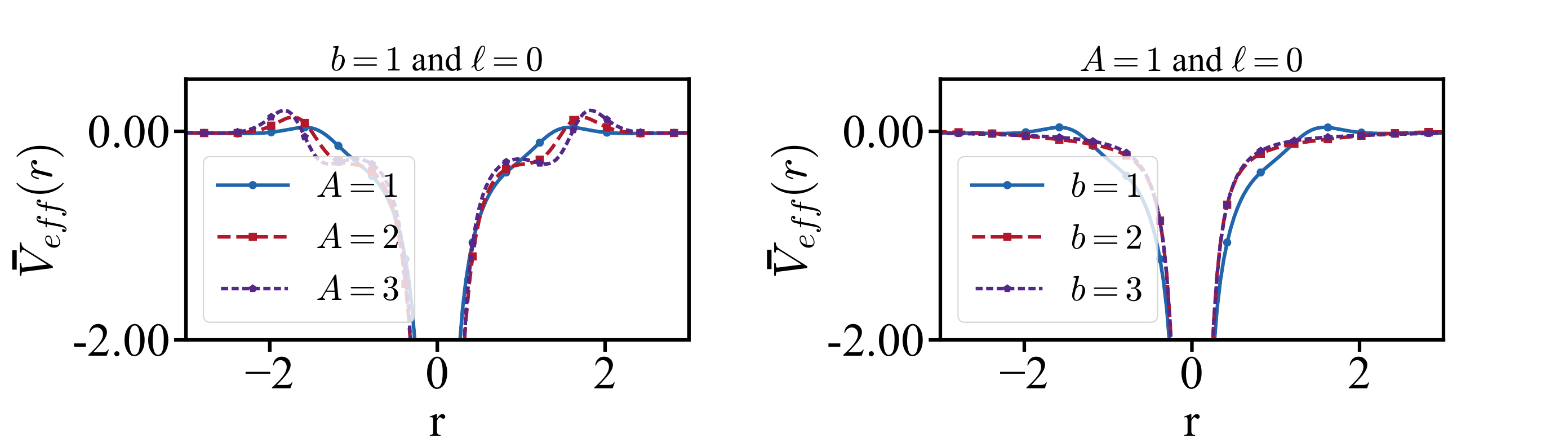}
    \caption{Plots of the effective potential $\bar{V}_{eff}$ for the electron on the quantum Gaussian ripple orbital angular momentum $\ell=0$. In the left panel we have considered $b=1$ and three values of $A$, namely $A=1$, $A=2$ and $A=3$. In the right panel we have considered $A=1$ and three values of $b$, namely, $b=1$, $b=2$ and $b=3$.}
    \label{FIG2}
\end{figure*}

Before obtaining the bound states and their respective spectra, let us discuss some qualitative features of Eq. \eqref{effective_pot}. Taking into account the angular symmetry of the ripple, the effective potential has the shape, in three dimensions, of a funnel of depth minus infinity. The Fig.\ref{FIG2} represents a cross-section of this surface, for $\ell=0$. Therefore, the domain of the figure is presented from minus infinity to plus infinity. The Fig. \ref{FIG2}a shows the effective potential, $\bar{V}_{eff}(r)$, for the of set parameters $b=1$ and three values of $A$, namely, $A=1$, $A=2$ and $A=3$. The Fig. \ref{FIG2}b is for the set of parameters $A=1$ andthree values of $b$, namely, $b=1$, $b=2$ and $b=3$. In general, $\bar{V}_{eff}(r)$ exhibits a symmetric potential well centered at the peak of the Gaussian-like ripple, i.e., at $r=0$. In the limit $r\rightarrow0$, the effective potential $\bar{V}_{eff}(r)$ goes to minus infinity, i.e., $\lim_{r\rightarrow 0}\bar{V}_{eff}(r)=-\infty$. From Fig. \ref{FIG2}a, We observed that the increase in the geometric parameter $A$ produces oscillations in the potential at the edge of the well, which corresponds to the region at the base of the ripple. In the next section, we will discuss how these oscillations can produce bound states, in this system. In \ref{FIG2}b, We observe that by increasing the geometric parameter $b$, the oscillations at the edge of the potential are smoothed.  

\begin{figure*}
    \centering
    \includegraphics[scale=0.4]{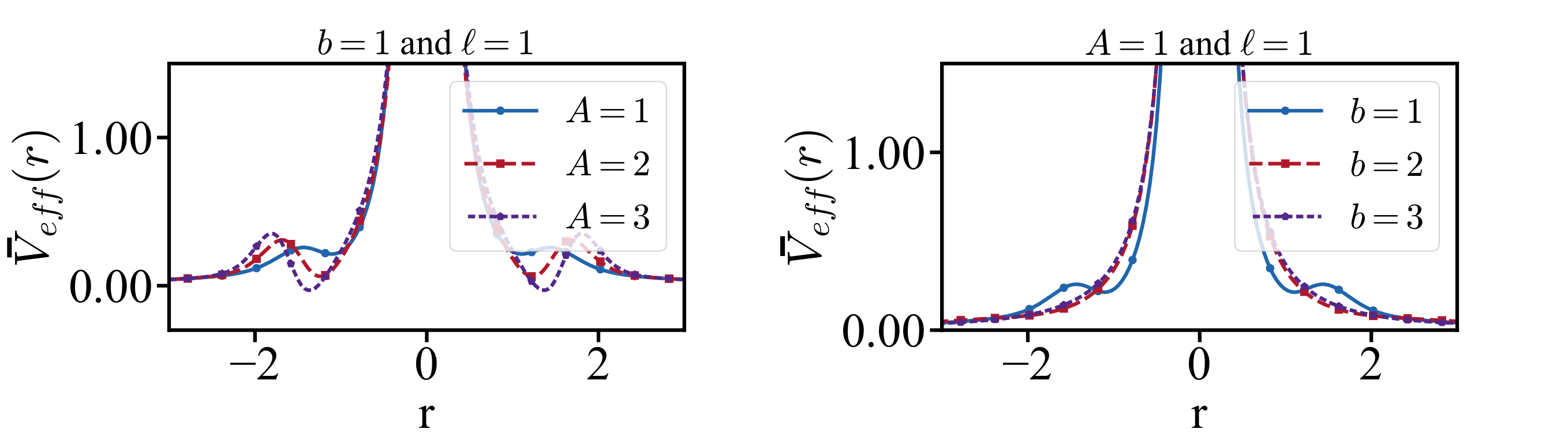}
    \caption{Plots of the effective potential $\bar{V}_{eff}$ for the electron on the quantum Gaussian ripple orbital angular momentum $\ell=1$. In the left panel we have considered $b=1$ and three values of $A$, namely $A=1$, $A=2$ and $A=3$. In the right panel we have considered $A=1$ and three values of $b$, namely, $b=1$, $b=2$ and $b=3$.}
    \label{FIG6}
\end{figure*}

Fig. \ref{FIG6} shows the general behaviour of the effective potential $\bar{V}_{eff}(r)$ for several values of the parameters $A$ and $b$ in the case where $\ell=1$. When the orbital angular momentum is taken into account the potential well gives way to a infinite potential peak centered at $r=0$. In Fig. \ref{FIG6}a, we have set $b=1$ and considered $A=1$, $A=2$ and $A=3$. As the geometric parameter $A$ increases, oscillations in potential also appear, but at the base of this potential peak.
In Fig. \ref{FIG6}b, we observe that the potential oscillations at the base of the peak are smoothed as the geometric parameter $b$ increases.

\section{Bound states}\label{section333}

In the case with no orbital angular momentum, that is, when $\ell=0$, our system can present some bound states. The aim of this section is therefore to investigate how the Gaussian-like ripple parameters $A$ and $b$ affect this bound states of the system.

\begin{figure*}
    \centering
    \includegraphics[scale=0.25]{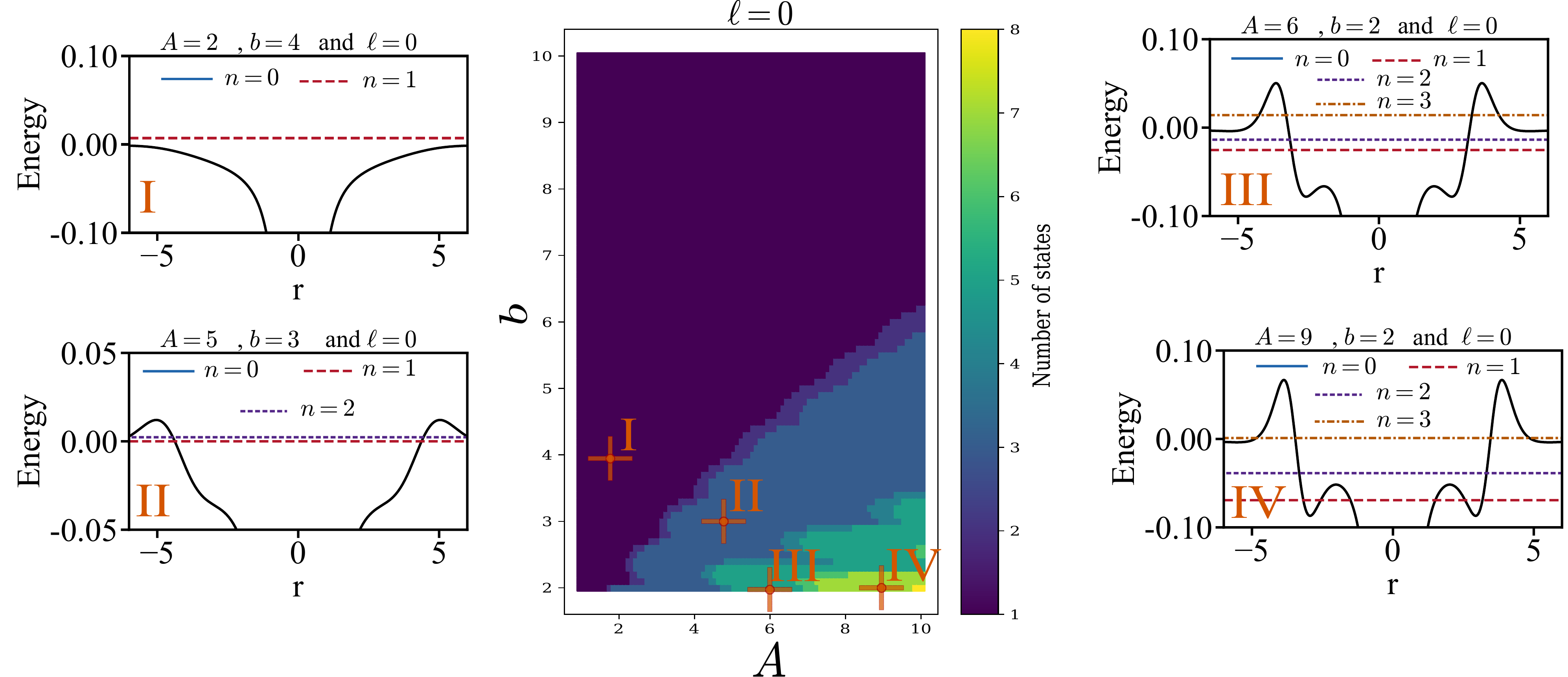}
    \caption{Bound states for the Gaussian-like ripple. The central color map shows us how the parameters $A$ and $b$ affect the number of bound states of the system.}
    \label{fig12}
\end{figure*}

First, let us address the question of the number of bound states present in a phase space spanned by the geometric parameters $A$ and $b$, say, $(A,b)$-space. In Fig. \ref{fig12}, we find a color map that shows the number of bound states related to the $(A,b)$-space. It becomes clear that, for a fixed value of $b$, the number of bound states increases as we increase the value of $A$. On the other hand, the number of bound states decreases when we fix the value of $A$ and increase the value of $b$, emphasizing the fact that the confined states of the system are highly dependent on the geometry. In Fig. \ref{fig12}, we have also highlighted four configurations of interest: I, II, III, and IV. For each of these configurations, we have depicted the effective potential $\bar{V}_{eff}(r)$ along with the bound states of the system. 

In configuration I, where $A = 2$ and $b = 4$ (see top left
panel), only one bound state is observed. This corresponds to the ground state $n = 0$, which is located precisely at the peak of the ripple, considering that in this region we have a well of infinite potential. The state $n = 1$, as we can see, is propagating. 

In configuration II, where $A=5$ and $b=3$ (see bottom left panel), three confined states are observed: the fundamental state $n=0$ at the bottom of the potential well, and two weakly confined states corresponding to the first and second excited states $n=1$ and $n=2$, respectively. Note that the latter are almost degenerate. These states are located at the edge of the potential well, that correspond to the ripple base.

In configuration III, where $A=6$ and $b=2$ (see top right panel), the system exhibit six bound states. We have plotted just three of them, namely, $n=1$, $n=2$, and $n=3$. Note that the gaps $\Delta_{12}=E_2-E_1$ and $\Delta_{23}=E_3-E_2$ are explicitly different, but not in a dramatic way. This fact could be related to an anharmonicity factor that may lead to effective qubit encoding in the energy levels of the system, adhering to (at least some of) the DiVincenzo criteria. 

Finally, in configuration IV, where $A=9$ and $b=2$ (see bottom right panel), the system exhibit eight confined states. As before, we have depicted only three of them, namely, the first three excited states $n=1$, $n=2$, and $n=3$. Similarly to the configuration III, we have a slight difference between $\Delta_{12}$ and $\Delta_{23}$. However, the difference between the gaps becomes smaller. When this difference is too small, a qubit encoding becomes impossible due to issues related to control and state addressing. All indications suggest that there must be an optimal geometric configuration of the ripple to ensure qubit encoding.

\section{Final Remarks}\label{section444}

In this paper, we investigate how geometry and orbital angular momentum affect the electronic states of an electron confined to a quantum ripple surface. We consider that the system’s dynamics are described by a Schrödinger-type equation in the presence of a confining potential known as the Costa potential. Through an appropriate parametrization for a Gaussian-type ripple surface, we derive all geometric quantities of interest, such as the interval, the metric, and the Christoffel symbols. The Costa potential $V_{dC}$, which depends on the mean and Gaussian curvatures, was also calculated, where it was found to assume only negative values for $r\neq 0$, regardless of the geometric parameters $A$ and $b$ that characterize the corrugation. At $r=0$ and as $r\rightarrow 0$, $V_{dC}$ assumes a value of zero. Additionally, we found that the Costa potential exhibits a global minimum whose depth increases with increasing parameter $A$ and decreases with increasing parameter $b$.

The axial symmetry of our problem also allowed the use of the method of separation of variables for partial differential equations, which, after some transformations on the separable solution that depends solely on the coordinate $r$, led to a Schrödinger-type equation describing the dynamics of a position-dependent mass system $M(r)$ in the presence of an effective potential $\bar{V}_{eff}$. The behavior of $M(r)$ with $r$ was analyzed, where it was found that its maximum value is always reached at the same point regardless of the parameter $A$, but tends to shift to the right as the value of $b$ increases, since $r_{max}=b/\sqrt{2}$. From our analysis, we also concluded that $M(r)$ can be written in terms of the Gaussian curvature $K$, further emphasizing that the system is highly dependent on the geometry presented here. Regarding the effective potential $\bar{V}_{eff}$, we analyzed it in four cases. The first two are related to the absence of orbital angular momentum, i.e., when $\ell=0$. In general, $\bar{V}_{eff}$ exhibits a very sharp symmetric potential well centered at the peak of the ripple. How sharp this well will be depends directly on the values assumed by the geometric parameters $A$ and $b$, that produces oscillations at the edge of this well of infinite potential. In the last two cases, we assumed $\ell=1$ and, unlike the previous two, the effective potential appears as an infinite peak centered at the origin of the ripple. Here infinite potential barrier depends directly on the values assumed by the parameters $A$ and $b$. 

Finally, we investigated how the geometric parameters $A$ and $b$ affect the bound states of the system in the absence of orbital angular momentum. A color map showing the number of bound states present in the phase space $(A,b)$ was presented. We concluded that for fixed values of $b$, the number of bound states increases with increasing $A$. Conversely, the number of bound states decreases when we increase the value of $b$ and fix $A$. This indicates that the confinement is highly dependent on the geometry presented here. At the end of Section III, we also discussed four distinct configurations of the phase space $(A,b)$, where we concluded that the ground state $n=0$ is always highly confined. We highlight here that in configurations with more than two bound states, we observed energy gaps that are explicitly different from each other, which may be associated with an anharmonicity factor that can lead to effective qubit encodings in the system’s energy levels. However, the difference between the gaps depends on the specific phase space, which may become small enough to lead to control and state addressing problems, thus preventing qubit encoding. Our work suggests that there should be an optimal geometric configuration of the ripple to ensure qubit encoding.

\section{Acknowledgments*}

JF would like to thank Alexandra Elbakyan and Sci-Hub, for removing all barriers in the way of science and the Funda\c{c}\~{a}o Cearense de Apoio ao Desenvolvimento Cient\'{i}fico e Tecnol\'{o}gico (FUNCAP) under the grant PRONEM PNE0112- 00085.01.00/16 and the Conselho Nacional de Desenvolvimento Científico e Tecnol\'{o}gico (CNPq) under the grant 304485/2023-3.


\end{document}